\documentclass[11pt,a4paper]{article}
\pdfoutput=1
\usepackage{lineno}

\usepackage[utf8]{inputenc}
\usepackage[T1,T2A]{fontenc}
\usepackage{nameref,hyperref}

\usepackage{amsmath}

\usepackage{csquotes}
\usepackage{subcaption}

\usepackage[disable]{todonotes}
\usepackage{authblk}

\usepackage[backend = biber, bibencoding=utf8]{biblatex}
\begin{document}

\title{Statistical time analysis for regular events with high count rate}

\author{Alexander Nozik}

\affil{Institute for Nuclear Research RAS, prospekt 60-letiya Oktyabrya 7a, Moscow 117312, Russian Federation}
\affil{	Moscow Institute of Physics and Technology, 9 Institutskiy per., Dolgoprudny, Moscow Region, 141700, Russian Federation}

\date{}
\maketitle

\begin{abstract}
    In physics it is frequently needed to precisely measure the count rate of some process. Quite often one needs to account for electronics dead time, pile-up and other features of data acquisition system to avoid systematic shifts of the count rate. In this article we present a statistical mechanism to diminish or completely eliminate systematic errors arising from the correlation between the events. Also we present examples of application of this method to the analysis of \textquote{Troitsk nu-mass} and \textquote{Tristan in Troitsk} experiments.
\end{abstract}

%
\section{Introduction}

One of the frequent types of measurements in physics is count rate measurement, where there are some events with the constant rate and one needs to measure this rate precisely. Usually it is done by simply dividing the total number of events in a time frame by the length of this time frame: $\mu = N_f / T_f$. In this case the time information from the events is discarded. There are several problems which could affect those measurements:
\begin{itemize}
    \item Dead time and/or event correlations for higher count rate. If events are too close to each other, some of them could not be counted, or the efficiency of detection could be reduced.
    \item Correction for the frame length for a smaller count rate. If the time between events is comparable to the length of a measurement interval, one could get an incorrect estimate of the count rate due to additional time between the first event in a frame and after the last event.
    \item Irregular background events, which do not have constant rate, but occur in short bursts, randomly distorting the measured count rate.
\end{itemize}

The standard technique to deal with those problems is to plot a histogram of time between subsequent events and qualitatively investigate the deviation from exponential law. One can also try to fit the histogram with exponential and extract the value of the count rate, but such a method brings additional systematic errors due to histogram binning. The statistical analysis technique, presented in this article, allows to extract full information from the time distribution of events (based on time window selection rule), search for anomalies in the distribution and correct those anomalies.

The technique was used to analyze the data of \textquote{Troitsk nu-mass} experiment in search for sterile neutrino in beta-decay (\cite{Abdurashitov:2015jha}). In this experiment we deal with both small count rate (about few Hz near the endpoint of the beta-decay spectrum) and relatively high count rate (up to 20-30 kHz). The major problem is the electronics dead time of about 6.5 $\mu s$ which could not be measured with sufficient precision and produces a major contribution to the systematic error. Also, it was discovered earlier during the previous experiment(\cite{Aseev:2011dq}), that there are short irregular bursts of background events. Those bursts do not affect sterile neutrino search, but still could be studied and eliminated. Currently, the main detector of \textquote{Troitsk nu-mass} is replaced by new high-speed segmented \textquote{TRISTAN} detector prototype (\cite{BRUNST2018}), which has smaller dead time and smaller count rate per channel (the project is called \textquote{Tristan in Troitsk}). The effect of dead time in this setup is much smaller but still must be investigated and accounted for.

There are a lot of articles about correct dead time accounting in different experiments. Usually they present  Monte-Carlo simulations of the dead time and correlation effects (a Monte-Carlo simulation was also done during the preparation of this work) or analytic study of statistical properties (like \cite{Kornilov:14}). \cite{Pomm__2015} gives a great overview of different ways to account for dead time in nuclear measurements, but does not discuss cases where the dead time is not exactly known. Also there are a lot of works about experimental estimation of dead time with different methods (like Fourier analysis), but the problem of unknown exact dead time and/or correlation effects seems to be rather rare and mostly missed by the community.

\section{Statistical time analysis}
\label{sec:stat}

Consider a simple Poisson process: independent events coming at a constant rate. The distribution of time intervals between events follows exponential distribution:
\begin{equation}
    \label{eq:exp}
	p(t) = \mu e^{-t \mu},
\end{equation}
where $\mu$ is the count rate and and $\tau = 1/\mu$ is the average time between events. In case events strictly follow this distribution, one could extract the count rate by maximizing likelihood function:
\begin{equation}
    \label{eq:likelihood}
    L(\mu)=\prod_{i=1}^N{p(t_i)}=\mu^{N} \exp \left( -\mu \sum_{i=1}^N{t_{i}} \right),
\end{equation}
where $N + 1$ is the total number of events and $t_i$ is a time distance between event number $i$ and $i+1$. Taking logarithm of (\ref{eq:likelihood}) and differentiating the result over $\mu$ one gets:

\begin{equation}
    \frac{\partial \ln L}{\partial \mu} = - \mu N + \mu^2 \sum_{i=1}^N{t_{i}}
\end{equation}

Equating this derivative to zero, one gets the solution:
\begin{equation}
    \label{eq:simple}
    \mu = \frac{N}{\sum_{i=1}^N{t_{i}}}.
\end{equation}

Designating $\sum_{i=1}^N{t_{i}}$ as $T$, one gets familiar expression $\mu = N/T$. While this solution coincides with the obvious definition of count rate as number of events in allotted time, there is a minor nuance. In general, one takes all events and the total measurement time, and in this case one takes all but one event and total time \textbf{between} the first and the last event, ignoring the time before the first event and after the last one. This difference is irrelevant for high count rates, but could matter for extremely low count rates when times between events is comparable with total measurement time.

Now let us suppose that there is a distortion of time distribution, which affects small time delays. Typical cases of such distortions are:
\begin{itemize}
    \item electronics dead-time;
    \item after-pulses, positive and negative event shape tails and other effects which could introduce correlation between nearby events;
    \item short-time high frequency noise bursts.
\end{itemize}

The dead time is usually taken into account when calculating the total count rate, but in cases when average distance between events is compatible with dead time, errors introduced by the incorrect determination of the dead time could be significant. The problem is complicated by the fact that hardware dead time is not constant and depends on different parameters like signal amplitude (\cite{Abdurashitov:2015jha}). After-pulses and event correlations are easy to miss even when analyzing time distributions. Noise bursts play a significant role when one works with small signal to noise ratio (low signal count rate) and could be seen by the naked eye in event distribution, but could not be eliminated by simple means.

In order to exclude systematic error from those effects, we propose to use a time cutoff. Let us choose arbitrary time $t_0$ and filter the event chain to leave only events with a delay greater than $t_0$. In this case the shape of time distribution above $t_0$ will not change, but there will be a change in distribution normalization and (\ref{eq:exp}) will look like this:
\begin{equation}
    \label{eq:pstar}
    p^*(t) = \mu e^{t_0 \mu}
    \begin{cases}
        e^{-t \mu} & t \geq t_0 \\
        0 & t < t_0  \\
    \end{cases}.
\end{equation}

The likelihood in this case looks like:
\begin{equation}
    \label{eq:like-cutoff}
	L(\mu) = \prod_{i=1}^{N}{p(t_i)} =  \left( \mu e^{t_0 \mu} \right)^N 
    	\exp{\left( - \mu \sum_{i=1}^{N}{t_i}\right)},
\end{equation}

where $t_i$ are the experimental intervals between events greater then $t_0$ and $N$ is a total number of those intervals. The likelihood logarithm looks this way:
\begin{equation}
	\ln L(\mu) = N \ln \mu + \mu N t_0 - \mu T
\end{equation}

The maximum of $L(\mu)$ corresponds to:
\begin{equation}
	\label{eq:cutoff}
	\mu = \frac{1}{\frac{T}{N} - t_0}~or~\tau = \frac{T}{N} - t_0.
\end{equation}

The difference between uncut solution (\ref{eq:simple}) and (\ref{eq:cutoff}) is additional term $t_0$ in average time estimation. Using Gaussian approximation, one could also get an uncertainty for this estimate. The statistical uncertainty for $\mu$ is defined by the same formula as for regular one $\sigma_{\mu} / \mu = 1 / \sqrt{N}$. 

The estimate could be also obtained by grouping data in a histogram and fitting it, but that approach is much slower since it involves non-linear curve fit and introduces additional systematic error from grouping data into a histogram.

One important note about this analysis is that it does not make any additional assumptions about the signal beyond the fact that events with $t > t_0$ are statistically independent. It produces mathematically correct results for any count rate and any cutoff time. Of course, for large cutoffs, the loss of statistical sensitivity will be significant.

Another important remark concerns the selection process for rejected events. If one wants to filter some noise or unwanted events, the method does not guarantee that all filtered out events are \textquote{bad} and all saved events are \textquote{good}. Usually one expects the \textquote{bad} event to come after \textquote{good} one, but it is not necessary the case. The method could be run on the reversed event chain, where the time difference is calculated backwards. If one wants to reliably get some properties of a signal beyond simple count rate, one needs to be careful to compare the results with forward and backward chains and combine time of arrival analysis with other techniques like amplitude analysis.

The distribution (\ref{eq:pstar}) accounts only for the lower boundary of $t$ since most distortions occur for smaller times, but it could be modified to include the upper boundary (for example for short measurements with small count rates). Also it is possible to use Bayesian analysis techniques to apply any kind of prior information on $t$.

A similar concept was discussed in \cite{GILAD201753}. In that article authors also introduce an artificial cutoff time, but instead of completely ignoring data below the cutoff, they extrapolate the time distribution to the zero delay time. The statistical time analysis has a significant advantage because it does not rely on generally unstable interpolation techniques and produces mathematically correct statistical errors. Also the backward extrapolation method, presented in \cite{GILAD201753}, does not solve the problem of cutoff time choice and instead relies on visible features of time distribution.

\section{Cutoff scan}

A powerful technique that could be used in time analysis is the cutoff time scan. For a single set of data one can estimate count rate with different $t_0$ and make a plot of $\mu$ versus $t_0$.

\begin{figure}
    \centering
    \begin{minipage}[t]{0.48\textwidth}
        \centering
        \includegraphics[width = \textwidth]{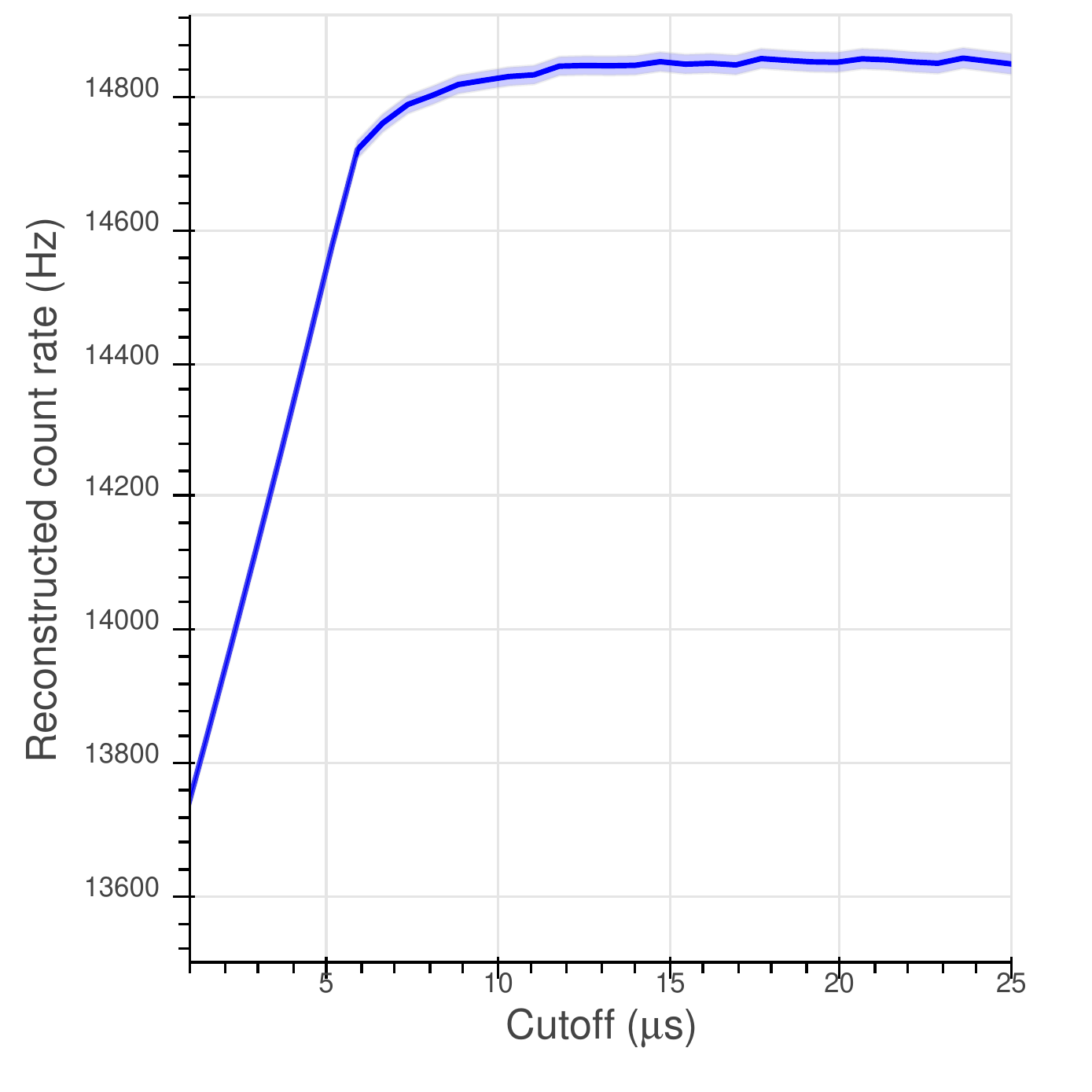}
        \caption{The cutoff scan for typical \textquote{Troitsk nu-mass} data at count rate about 15~kHz.}
        \label{fig:scan}
    \end{minipage}
    ~
    \begin{minipage}[t]{0.48\textwidth}
        \centering
        \includegraphics[width = \textwidth]{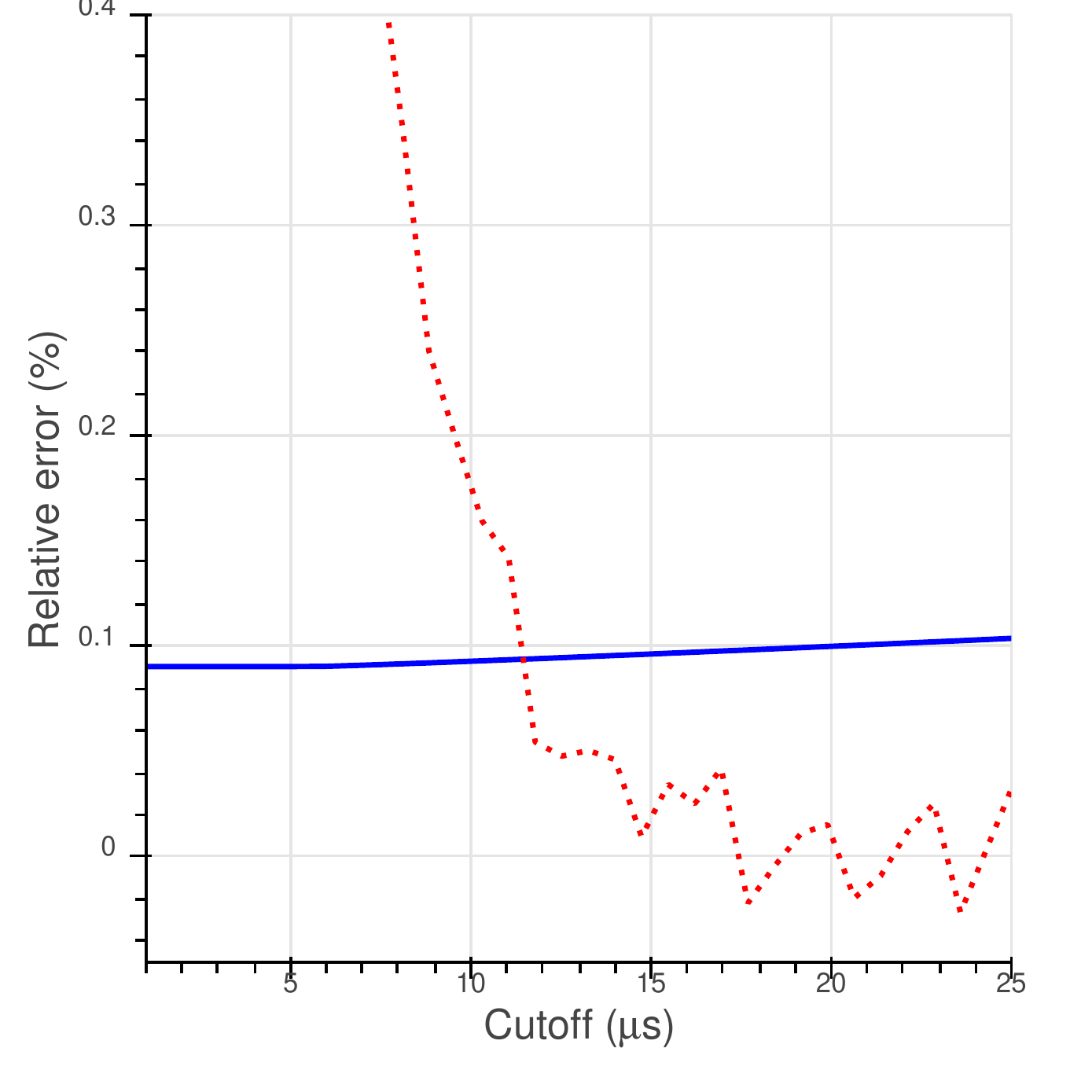}
        \caption{Solid line - relative increase of the statistical error with respect to the case where no cutoff is applied. Dotted line - relative difference between the computed count rate and the plateau value in Fig.~\ref{fig:scan}.}
        \label{fig:scan_err}
    \end{minipage}
\end{figure}

Fig.~\ref{fig:scan} shows cutoff scan for typical \textquote{Troitsk nu-mass} data. The figure shows data corresponding to 90 seconds of data acquisition, which is about 1\% of all data gathered in 2017 at this voltage point and of the order of 0.01\% of total acquisition time during that year (the count rate is different for different points). The larger samples are harder to interpret due to the effect discussed in section~\ref{sec:npn-uniform}. The greyed area around the curve represents the resulting statistical error for given $t_0$. The values for different cutoff times are strongly correlated because they are based on almost the same data, so the real difference between neighboring points is much less than that error. 

The sharp rise to the left is caused by the electronics dead time which approximately equals $6~\mu s$ for this run. The linear rise could be easily explained by the equation~\ref{eq:cutoff}. Since there are no events in the time region between $t_0 = 0$ and $t_0 \approx 6 \mu s$, the $T/N$ term of the equation does not change and the change in $\mu$ looks like $\mu = 1/(T/N - t_0) \approx N/T (1 + t_0 N/T) $ which corresponds to linear function.

The region between $t_0 \approx 6 \mu s$ and  $t_0 \approx 15 \mu s$ demonstrates another feature, specific to \textquote{Troitsk nu-mass} data which in fact was found only after application of the cutoff scan procedure. This region is further discussed in section~\ref{sec:correlation}. The plot allows not only to find anomalies (significant deviation of dependence from constant) and establish $t_0$ that should be used, but also estimate the systematic shift caused by those anomalies and an increase of statistical error for different cutoffs.

Estimation of $t_0$ for final analysis based on a region of stability is fair from statistical point of view, since any choice of $t_0$ gives mathematically correct result, but one should be careful to use the same $t_0$ for all data sets. Choosing different $t_0$ for different data sets or using criterion not based on stability (for example selecting cutoff which gives the smallest count rate) could add additional information to the analysis.

\section{Dead time correction}

The dead time uncertainty is often a major source of systematic error. The problem comes from the fact that the dead time could not be experimentally estimated with sufficient precision, may change from one run to another and could depend on the event amplitude (all those problems are observed at \textquote{Troitsk nu-mass}). In order to avoid uncertainties from dead time estimation, one could select a $t_0$ cutoff slightly above the hardware dead time and estimate $\mu$ from (\ref{eq:cutoff}) using modified distribution (\ref{eq:pstar}). 

Since $t_0$ is set manually, it could be selected with any precision. So the only systematic limitation of the method (assuming that there are no correlations between the event with a time above $t_0$) is the precision of time measurement for an individual event. Of course statistical error will be slightly increased because some events with delay below $t_0$ are excluded from the analysis. The relative increase to statistical error due to event cut equals the square root of ratio between total number of events before and after the cut. The optimal choice of $t_0$ is different for each specific task. In general one wants to avoid loss of statistic sensitivity and select $t_0$ as close to the dead time as possible. Usually dead time is not exactly the same for all events (for example it depends on the event amplitude), so one should carefully examine time cutoff scan before selecting a cutoff for final analysis. The cutoff scan allows to observe the results of the choice. The increase in statistical error is defined by the enlarging error band, while systematical shift could be observed as a significant (well outside statistical band) deviation of reconstructed count rate from constant. One should note that if we have ideal case of non-extending dead time and set $t_0$ exactly to that value, we get the best possible statistical precision (not only for method, presented in this article, but also for all other possible methods) since we use all information from the data (Cramer–Rao bound).

The systematic error from time measurement precision could be estimated as a difference in reconstructed count rates in a range of $t_0$ corresponding to event time measurement uncertainty. it will depend on the curvature of the cutoff scan plot and obviously almost zero for flat regions.

\section{Correlation analysis}
\label{sec:correlation}

Another problem that arises is the correlation between events. For example, Fig.~\ref{fig:scan} shows not only a sharp fall below the dead time, but also a smooth increase from 6 up to 15 $\mu s$, which has different nature. The shape of the signal from the\textquote{Troitsk nu-mass} detector is demonstrated in \cite{Chernov:2018ysc}. The problem is that if the second signal comes within $ 15~\mu s$ from the first signal, it falls on the \textquote{tail} of the previous signal and its amplitude is effectively diminished. Since there is a finite lower detection threshold, some events, that could be otherwise registered, fall below this threshold and are ignored, therefore producing lowered detection efficiency in the given frame. The magnitude of this effect depends on the count rate, therefore it is crucial for the\textquote{Troitsk nu-mass} analysis. One should note, that this effect alone would be important enough to use the cutoff scan even without uncertainty in dead time. 

The solid line at Fig.~\ref{fig:scan_err} demonstrates the rise of statistical error (the same data as for Fig.~\ref{fig:scan}).The dotted line shows the relative systematic shift of reconstructed count rate assuming the plateau on Fig.\ref{fig:scan} corresponds to the real count rate (the plateau value in this case is calculated as average in the $t_0$ region from 17 to 25 $\mu s$). One should note that the solid line shows statistical error for a single data set, so for the data, combined from multiple data sets, the error will be diminished (we usually independently analyze data blocks approximately 5-10 times larger then the data sample on the figure and then combine the fit results without combining the data itself). The dotted line corresponds to at least partially systematic error: the oscillations for larger cutoffs will decrease with statistics, but the rise closer to zero is persistent. In case of \textquote{Troitsk nu-mass} the total result is systematics-limited so we tend to use a conservative cutoff of $15 \mu s$. Currently, we managed to mitigate the statistics loss by using advanced pulse separation algorithm described in \cite{dubna_time} and \cite{Chernov:2018ysc}. It allows to lower effective dead time to $\sim 3~\mu s$ without introducing anomalies in the cutoff scan picture.

The cutoff time scan technique does not make any assumptions on the nature of correlations. On the contrary, anomalies in the cutoff scan could help to find specific problems like it was done in \textquote{Troitsk nu-mass} analysis. If the anomaly is discovered and there is no way to correct or understand its origin, the only way to treat the problem is to ignore anomalous events and use only \textquote{good} ones, which is exactly what we propose in this article. If there is a way to treat the problem (for example use different readout and correction method), then time-scan could be used as a quality control procedure to check that those methods do not produce additional anomalies (this quality control is currently used in \textquote{Troitsk nu-mass} as a part of automated analysis procedure). 

Some previous works (like \cite{Schmidt1984}) try to treat different special cases of event correlations like multiple decay chains. The procedure explained in section~\ref{sec:stat} heavily relies on the assumption that the time distribution is exponential (at least for large values of $t$). It could not be applied in its current form to different cases like bi-exponential distribution. It still could be used if the distortions from the second exponent appear only for rather small delay times, but its effectiveness (in terms of statistical precision) will be lower than effectiveness of specialized techniques. In any case, a cutoff scan plot could be used as an auxiliary procedure to get the information about both expected and unexpected deviations from the exponential distribution law.

It is also possible to incorporate the additional information and understanding of the specific physical process by using more sophisticated prior distribution transforming (\ref{eq:exp}) into (\ref{eq:pstar}) and use \textquote{soft} statistical accept-reject procedure instead of straight time cut, but of course in this case simple analytic formulae won't work and the procedure should be prepared and studied for each specific case.

\section{Bunch noise rejection}

Time of arrival could be used not only for high count rate, but also for cleaning irregular background in low count rate part of the spectrum. In \textquote{Troitsk nu-mass} there are two sources of such irregular background:
\begin{itemize}
    \item Spectrometer electrode discharge. Micro-discharges produce very short (few milliseconds) high frequency bursts.
    \item Electron trapping in the spectrometer. Electrons born inside the main spectrometer sometimes become trapped inside between two magnetic mirrors. In this case the electron loses energy by ionizing residual gas in the spectrometer and in that process produces secondary electrons which could hit the detector. Those events look like long (few seconds) \textquote{bunches} of events with a slightly increased count rate.
\end{itemize}

Previously such noise was treated with a sliding window algorithm, which cut the whole time frames with count rate greatly exceeding the average count rate in the data point. The sliding window procedure was explained in \cite{Zadorozhniy} (in Russian) and later improved by introducing automatically adjusting window in \cite{Nozik:bunches} (also in Russian). Algorithms like this have a number of problems:
\begin{itemize}
    \item One needs to manually set the threshold value for a count rate. It could be calculated based on fixed probability value so the probability to cut the frame in the sample without noise is always the same (does not depend on the count rate). But still, this probability is defined manually and it is hard to estimate without a lot of simulations.
    \item The systematic error introduced by the procedure is hard to estimate without simulation.
    \item The effectiveness of filtering strongly depends on the ratio between noise rate and measured count rate. The method does not work for the average count rate more than a few Hz.
    \item The effectiveness of filtering depends on the correct guess of the frame length, because short frames are not effective for long bunches and long ones do not work well for short bunches.
\end{itemize}

\begin{figure}
    \centering
    \begin{subfigure}[b]{0.48\textwidth}
        \includegraphics[width=\textwidth]{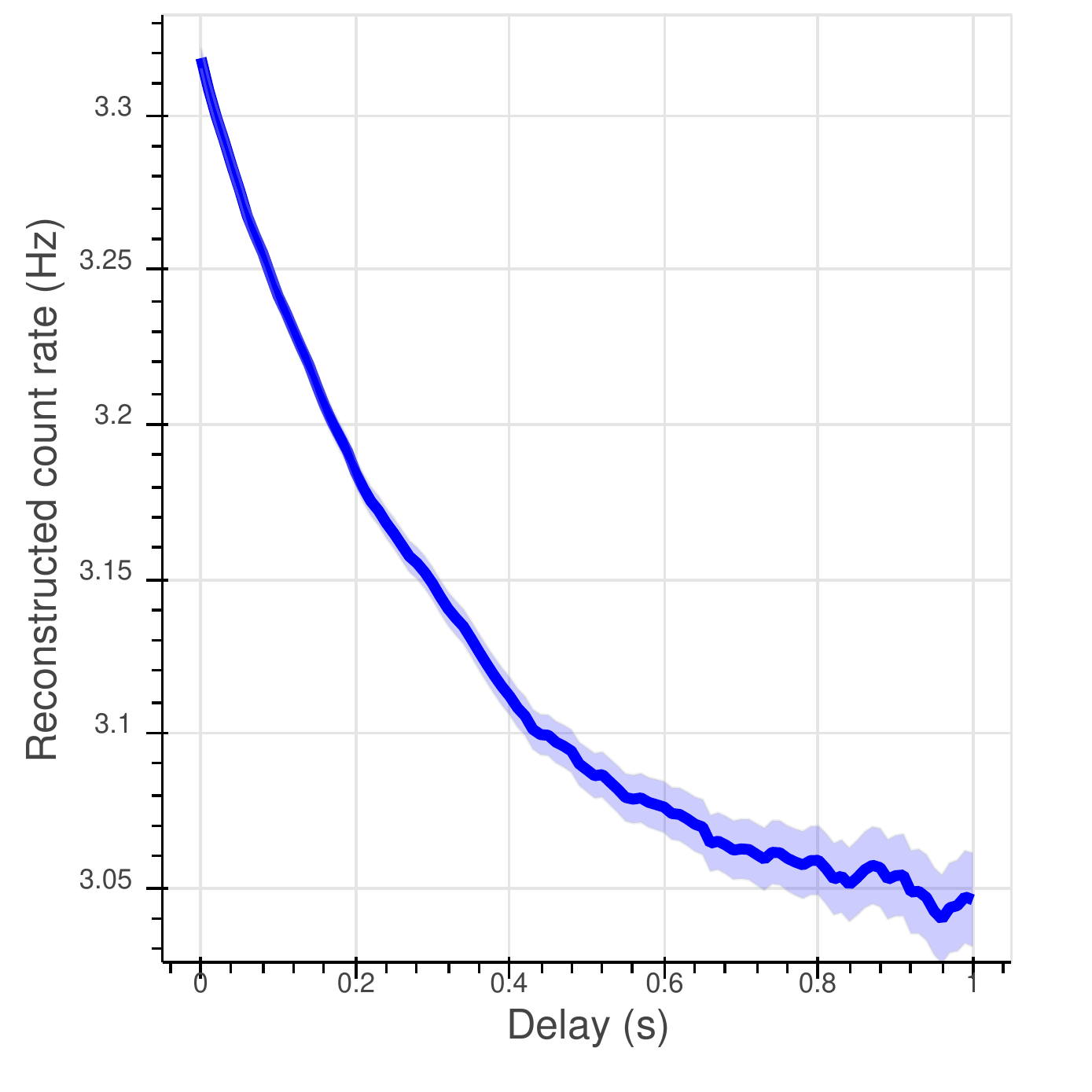}
        \caption{Count rate in bunch: 3 Hz.}
        \label{fig:bunches-3}
    \end{subfigure}
    ~
    \begin{subfigure}[b]{0.48\textwidth}
        \includegraphics[width=\textwidth]{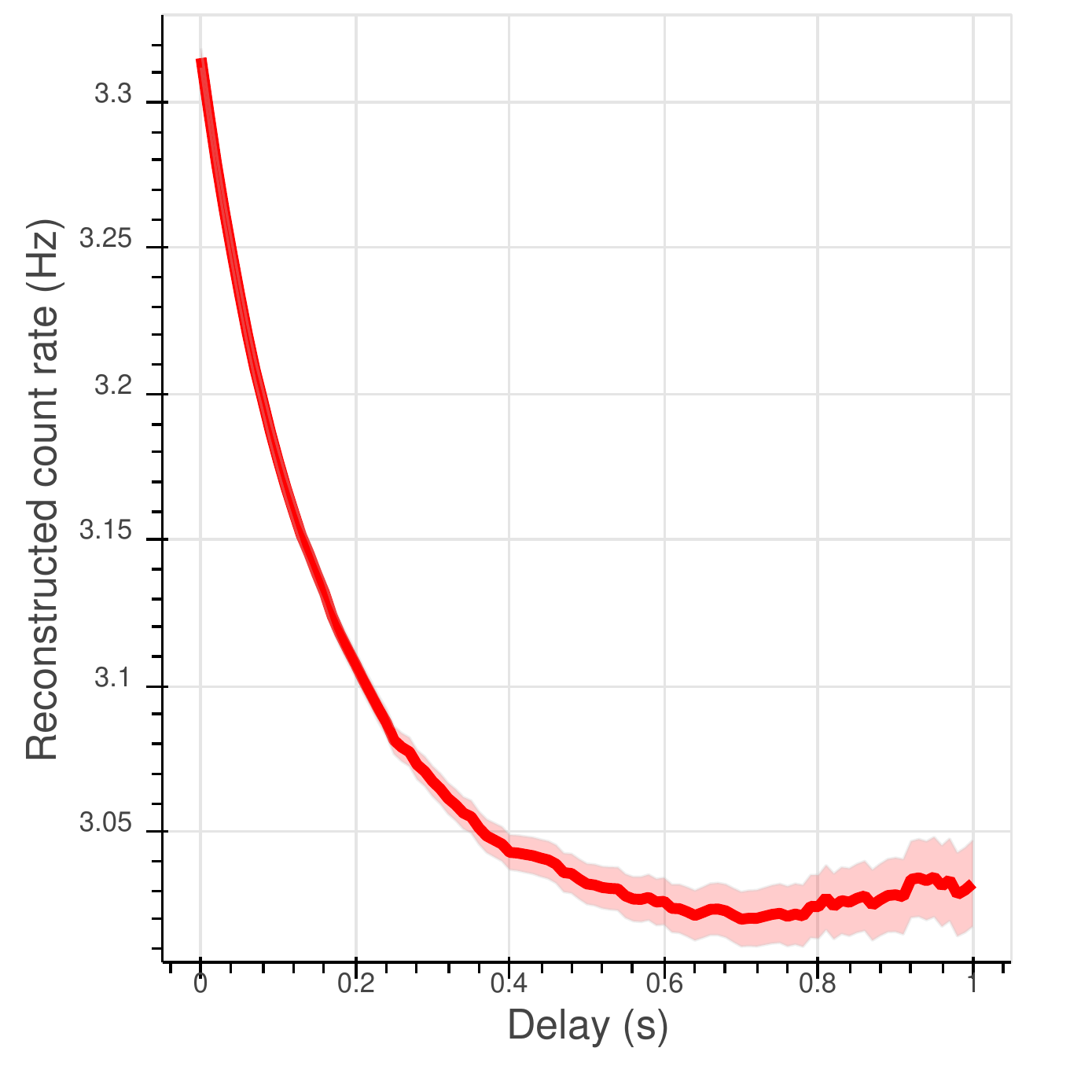}
        \caption{Count rate in bunch: 6 Hz.}
        \label{fig:bunches-6}
    \end{subfigure}
    \caption{A reconstructed count rate dependence on $t_0$ for count rate of 3Hz}
    \label{fig:bunches}
\end{figure}

The statistical approach allows to solve all those problems. In order to demonstrate a solution, let us use the simulation data with parameters similar to the ones observed at experiment:
\begin{itemize}
    \item The basic count rate (without bunches) --- 3~Hz. It is a typical count rate close to the beta-spectrum endpoint.
    \item The bunch length --- 5~s. The real trapped electron bunches usually have duration from 3 to 10~s. The discharge bunches have duration of about 0.1~s and are quite easy to discover with any method.
    \item For additional count rate of events in bunch, there are two cases: 3 and 6~Hz.
    \item The bunch rate is a few mHz so the probability of bunch overlap is extremely small and each bunch could be evaluated independently. The rate of bunches is adjusted for 3 and 6~Hz cases to make the total contribution from bunches the same in both cases. This was done to make plots easily comparable.
\end{itemize}
Fig.~\ref{fig:bunches} shows the cutoff scan plots for both bunch rates. 

It could be seen, that after selecting appropriate cutoff $t_0$, it is possible to mitigate bunch effect in both cases. In case of bunch count rate of $6~Hz$, the cutoff stabilization occurs for lower $t_0$ which allows to treat the problem with smaller loss of statistics. 

The $t_0$ parameter could be selected manually using a cutoff scan plot. Also in order to be completely unbiased one could choose a constant fraction of the events to be rejected $\gamma$ and adjust cutoff time for each actual count rate $r$ to approximately cut  this fraction: $t_0 = \frac{\ln(1-\gamma)}{r}$. In this case all data sets are treated exactly the same way. The effectiveness of count rate reconstruction in this case does not depend on bunch length.

The former approach with a sliding window introduced smaller statistical error overhead, but was not able to properly evaluate cases where count rate in a bunch was less or equal than the signal count rate (3Hz case) and gave significant offset when the actual bunch length deviated from the expected one. It was mitigated by running multiple scans with different window lengths which significantly complicated the analysis procedure. Also the running window algorithm did not work for higher base count rates and there was a critical count rate at which the algorithm was just switched off introducing possible kink. The cutoff scan could be used in conjunction with moving window algorithm to provide quality control after window filtration is done.

Currently, bunch noise rejection is not used in \textquote{Troitsk nu-mass} analysis since bunches do not affect sterile neutrino searches, so we do not study it in great detail in this article. But this method could be used for example in KATRIN experiment, which encounters similar problems (\cite{Arenz2018}).

\section{Non-uniform count rate case}\label{sec:npn-uniform}

\begin{figure}
    \centering
    \includegraphics[width = 0.9\textwidth]{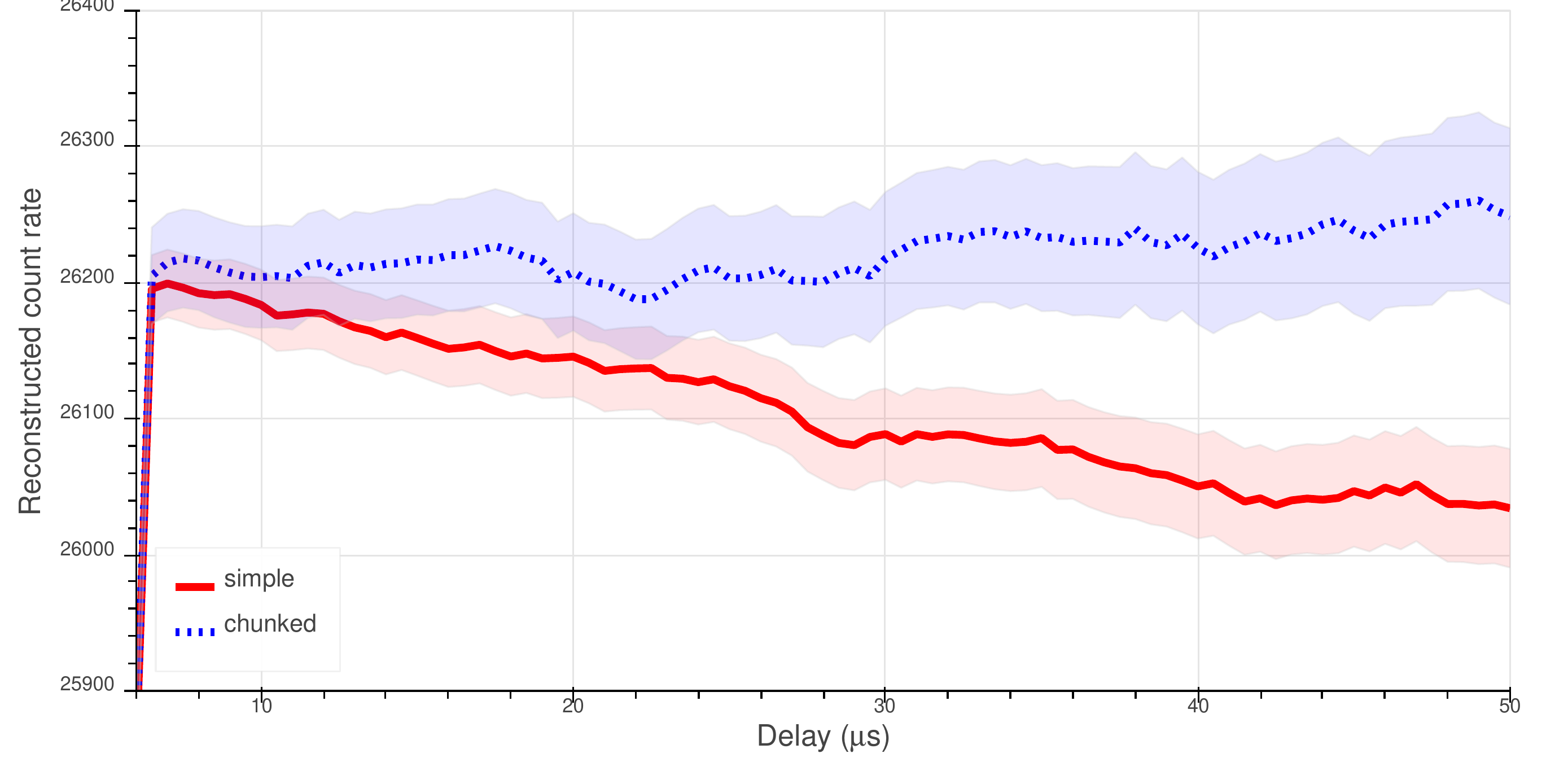}
    \caption{Analysis of changing count rate for three different cases: solid line shows the rate-cutoff dependency for simple analysis, dashed --- for weighted average of chunks of 5k events and dotted one --- for arithmetic mean with the same chunks. In all cases the same data was used. The error bands show the statistical error of resulting reconstructed count rate.}
    \label{fig:irregular}
\end{figure}

Even in a technically correct experiment, count rate is not always exactly the same. Consider a case of slowly changing count rate. For example in \textquote{Troitsk nu-mass} the drift of intensity could be up to $1\%$ during one measurement. In this case the distribution is not exactly exponential and the method could not be used as is. It could be solved by separating the event chain in small chunks (for example, 1000 subsequent events), calculating the count rate for each chunk and then averaging it. In this case count rate is the same during the chunk and the method works fine. 

Fig~\ref{fig:irregular} shows the result of simple cutoff scan for the whole data block and arithmetic mean of small chunks. The simulated data for this picture has initial count rate of 30~kHz which dropped by 25\% during 50-seconds measurement (the average count rate is 26.25 kHz). It could be seen, that simple splitting the chain in chunks improves the result, but using arithmetic mean solves the problem completely.

\section{Conclusion}

The histogram of distribution of times between events is commonly used to find irregularities in the events time distribution or to show lack of such irregularities, but the distribution is almost never used as a primary tool for analysis due to instabilities and loss of information caused by histogram fitting. In this work we presented the mathematically correct approach to extract the information about count rate directly from the time distribution (\ref{eq:pstar}) without grouping events in a histogram and thus without fitting procedure at all.

Additionally we presented a time difference cutoff scan --- the powerful technique, which could be used both for examining data for irregularities and to select the final cutoff time $t_0$ to cut those irregularities. The technique allows to work with a very high count rate without systematic effects and correctly evaluate statistical errors when dead time is present. Also it allows to perform systematic-free irregular noise filtering on small count rates and could be used with conjunction with other techniques as a quality control mechanism.

The approach very similar to the one used in this article could be found in \cite{meeks}. Instead of using a chain filter, authors propose to subtract the dead time from each of intervals and leave only those with a positive time. The resulting formula for count rate is equivalent to (\ref{eq:cutoff}), but the technique presented in this article is more concise and could be adjusted to more complicated priors.

In case of \textquote{Troitsk nu-mass} time analysis allowed to completely avoid systematic error from dead time uncertainty, by sacrificing a minor portion of statistics. Also we used the cutoff scan technique to find and eliminate minor flaws in electronics operation.

The work was supported with RFBR grant \textquote{17-02-00361 A}. I would like to thank Vladislav Pantuev for manuscript revision and constructive criticism of the work.

\printbibliography
\end{document}